\documentclass{svjour2}                    % onecolumn
\smartqed  % flush right qed marks, e.g. at end of proof
\usepackage{graphicx,color,amsmath}

\begin{document}

\title{
Protocols for copying and proofreading in template-assisted polymerization
%\thanks{Grants or other notes
%about the article that should go on the front page should be
%placed here. General acknowledgments should be placed at the end of the article.}
}
%\subtitle{Do you have a subtitle?\\ If so, write it here}

%\titlerunning{Short form of title}        % if too long for running head

\author{Simone Pigolotti \and Pablo Sartori         
}

%\authorrunning{Short form of author list} % if too long for running head

\institute{
  S. Pigolotti \at
           Universitat Politecnica de Catalunya,\\
             Edif. GAIA, Rambla Sant Nebridi 22, \\
             08222 Terrassa, Barcelona, Spain.\\
              Tel.: +34  93 739 8573\\
              Fax: +34  93 739 8000\\
              \email{simone.pigolotti@upc.edu}            \\
\and
P. Sartori\at
              Max Planck Institute for the Physics of Complex
  Systems.\\ Noethnitzer Strasse 38 , 01187, Dresden,
  Germany.
%             \emph{Present address:} of F. Author  %  if needed
}

\date{Received: date / Accepted: date}
% The correct dates will be entered by the editor

\maketitle

\begin{abstract}
  We discuss how information encoded in a template polymer can be
  stochastically copied into a copy polymer. We consider four
  different stochastic copy protocols of increasing complexity,
  inspired by building blocks of the mRNA translation pathway. In the
  first protocol, monomer incorporation occurs in a single stochastic
  transition. We then move to a more elaborate protocol in which an
  intermediate step can be used for error correction. Finally, we
  discuss the operating regimes of two kinetic proofreading protocols:
  one in which proofreading acts from the final copying step, and one
  in which it acts from an intermediate step.  We review known results
  for these models and, in some cases, extend them to analyze all
  possible combinations of energetic and kinetic discrimination. We
  show that, in each of these protocols, only a limited number of
  these combinations leads to an improvement of the overall copying
  accuracy.
\end{abstract}

\bigskip


\section{Introduction}\label{sec:intro}

Copying of information constitutes one of the key operations that
biological systems need to perform. Information is
cyclically and reliably transferred from a template to a copy during
DNA duplication, its transcription into mRNA, and also during protein
synthesis. In all these examples, copies must be accurate reproductions
of the template, as copy errors may compromise biological functionality. 

Accuracy of copying is easy to estimate close to thermodynamic
equilibrium. Because of detailed balance, the probability of a copy
error $\eta$ is $\eta_{\rm eq}\sim \exp[-\beta (\Delta E^w-\Delta
E^r)]$, where $\Delta E^w - \Delta E^r$ is the energy difference
between a right and a wrong incorporated monomer and $\beta=(k_B
T)^{-1}$ is the inverse thermal energy. Already in the 50s, before
detailed molecular studies of copying reactions were possible, Pauling
pointed out that structural differences between some amino acids are
small, so that their binding energy difference can not possibly
account for the accuracy required in protein synthesis
\cite{pauling1958festschrift}. Physically, this implies that high
accuracy can only be achieved by producing the copies via a
far-from-equilibrium thermodynamic process.

More recent experimental studies discovered that, indeed, most
biological copying reactions are composed of many out-of-equilibrium
intermediate steps, that can be identified in detail in some
cases. %From a more general point of view, it is interesting to seek
%for common principles underlying copying accuracy.
An important hallmark in the understanding of such complex copying 
processes was the proposal by Hopfield
\cite{hopfield1974kinetic} and Ninio \cite{ninio1975kinetic} of the
mechanism of {\em kinetic proofreading}. In kinetic proofreading, a discriminatory enzymatic reaction is complemented by an irreversible,
non-discriminating step that brings the complex to a high-energy
state. Kinetic proofreading was shown to reduce the minimum error down to
the {\em square} of the equilibrium error, and further down to higher powers
if multiple proofreading steps are considered
\cite{freter1980proofreading}. This idea had a deep impact as a
paradigm of error correction in biochemical reactions and helped
understanding the mechanisms underlying the high fidelity of many
copying systems, including DNA duplication \cite{loeb1982fidelity},
translation \cite{loeb1982fidelity}, and other cellular tasks
requiring high specificity such as T-cells recognition
\cite{mckeithan1995kinetic,franccois2013phenotypic}. Recently, a
generalization of kinetic proofreading to arbitrarily wired networks
has been explored by Murugan, Huse and Leibler
\cite{murugan2012speed,murugan2014discriminatory}

A parallel approach to biological copying was pioneered by Charles
Bennett, when he proposed a model in which a copy polymer is elongated
to match a template \cite{bennett1979dissipation}   (see also
  section \ref{sec:results}). An important
  conceptual aspect
of Bennett's model is that it describes the copy of a whole polymer,
i.e. a process in which
a large number of monomers are sequentially copied. This is in contrast to 
  stoichiometric models, such as Hopfield's proofreading model, which
describe copy of a single monomer. In Bennett's model, each
copy event is simply modeled as a single-step process, and energy
differences between right and wrong matches are neglected. This
simplicity, together with the steady-state nature of the solution,
made the model particularly appropriate to analyze the issue of
irreversibility of the copying process. However, Bennett's work received
relatively little attention until recently, when a renewed interest in
the physics of biological information processing has led to further
analyses in this direction
\cite{andrieux2008nonequilibrium,esposito2010extracting}.

\begin{figure}
\begin{center}
\includegraphics{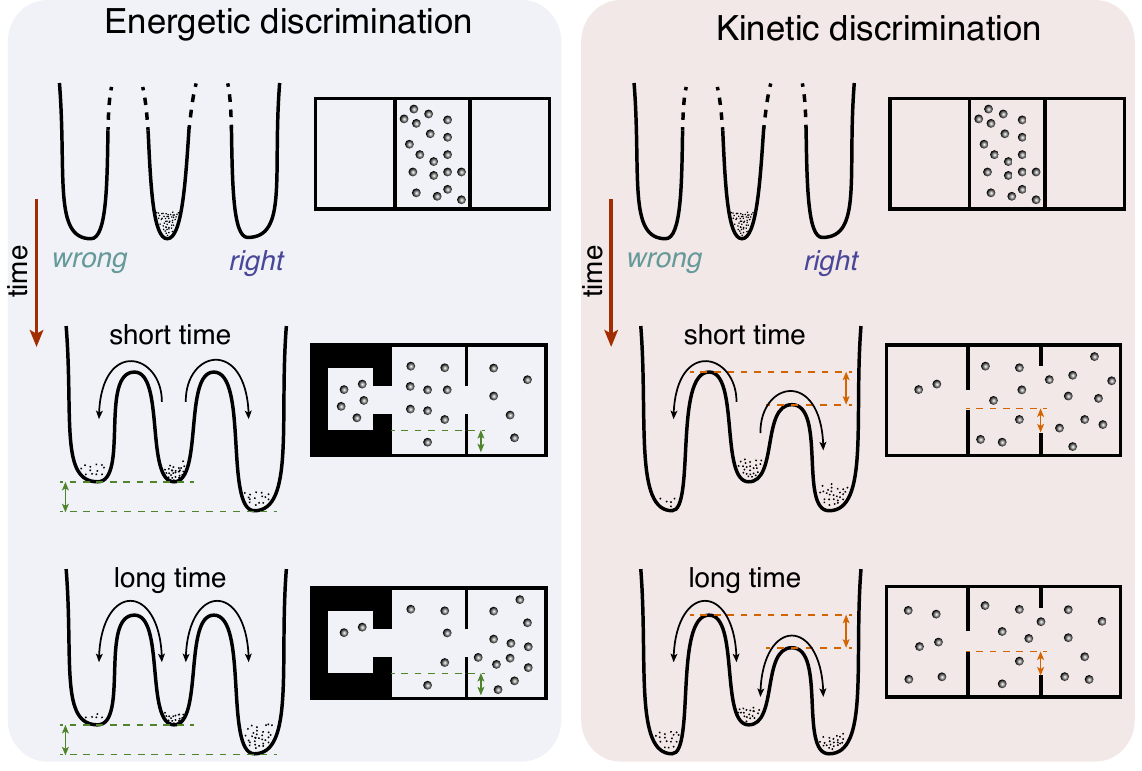}
\caption{{\bf Energetic and Kinetic discrimination.}
   In both panels (Energetic and Kinetic discrimination), the
    first column shows an energy landscape for a copy process. In
    particular, in the central (null) state, no monomer is bound to the
    copying machine. The left and right states represent respectively
    a bound monomer which do not match (wrong) or do match (right) the
    template. The second column presents an analogy in terms of a gas
    of diffusing particles, where the three states correspond to different
  chambers. In {\em Energetic discrimination}, the wrong and right
  states have different energy, while the kinetic barriers to reach
  them from the null state are the same. In the gas analogy, the
  openings leading to the
  right/wrong chambers have the same width but the two chambers have
  different volumes. If the system is prepared in the null state, at
  short times the probabilities of being in the right and wrong states
  are dictated by the energy barriers (opening widths) and are
  therefore approximately equal. At longer times, the system
  equilibrates and the error reduces the energy
  differences (chamber sizes). In {\em Kinetic discrimination}, 
the wrong and right
  states have the same energy, while the kinetic barriers to reach
  them from the null state differ. In the gas analogy, the
  openings leading to the
  right/wrong chambers have different widths but the two chambers have
  equal volumes. Preparing the system in the null state, at
  short times the probabilities of being in the right and wrong states
  are determined by the energy barriers (opening widths), leading to a
  low error probability. At longer times, the system
  equilibrates, so that the probability of being in the right/wrong
  states are the same.
  \label{fig:potential}}
\end{center}
\end{figure}

To understand a fundamental difference between equilibrium and
non-equilibrium copying, it is useful to introduce the concepts of
{\em energetic} and {\em kinetic} discrimination, see
Fig.~(\ref{fig:potential}). In both panels, these two concepts
are illustrated by means of two different conceptual examples. The first
is an energy landscape for a simple three-state copying model. 
The three states represent, from left to right, the state in which a
wrong monomer is bound, the one in which no monomer is bound, and the
one in which the right monomer is bound.  The second example is a
mechanical model in which the system is a gas of diffusing partices,
where the three chambers are the analogous of the three states of the
first example. In all cases, the system is initially prepared in the
null central state, top panes in Fig.~(\ref{fig:potential}). We are
interested in the evolution of the error, defined as the ratio of
probabilities of being in the wrong vs. right state as a function of time.

In energetic discrimination, the wrong and right states are separated
by the same kinetic barrier from the null state; however, the wrong
and right states have different energies. In the gas analogy, the
openings to reach the right and wrong chambers have the same width,
but the two chambers have different volumes. At short times, the error
is mostly determined by the kinetic barriers and is therefore
large. At longer times, the system equilibrates leading to a reduction
of the error. Conversely, in kinetic discrimination, the kinetic
barrier to reach the wrong state is higher than that to reach the
right state, but the right and wrong states are isoenergetic. In the
mechanical analogy, the openings leading to the right and wrong
chambers have different widths, but the two chambers have the same
volume. At short times, the error is determined by the kinetic barrier
difference and therefore is low. At longer times, the system
equilibrates and the error grows due to the lack of energy differences.

We recently discussed how kinetic and energetic discrimination are
mutually exclusive in single-step protocols \cite{sartori2013kinetic}:
it is impossible to improve accuracy at intermediate times by
combining their effects. Indeed, most models in the literature
consider only one form of discrimination at a time: for example,
Hopfield's model \cite{hopfield1974kinetic} does not consider kinetic
discrimination, while Bennett's model \cite{bennett1979dissipation}
does not consider energetic differences.

\begin{figure}
\begin{center}
\includegraphics{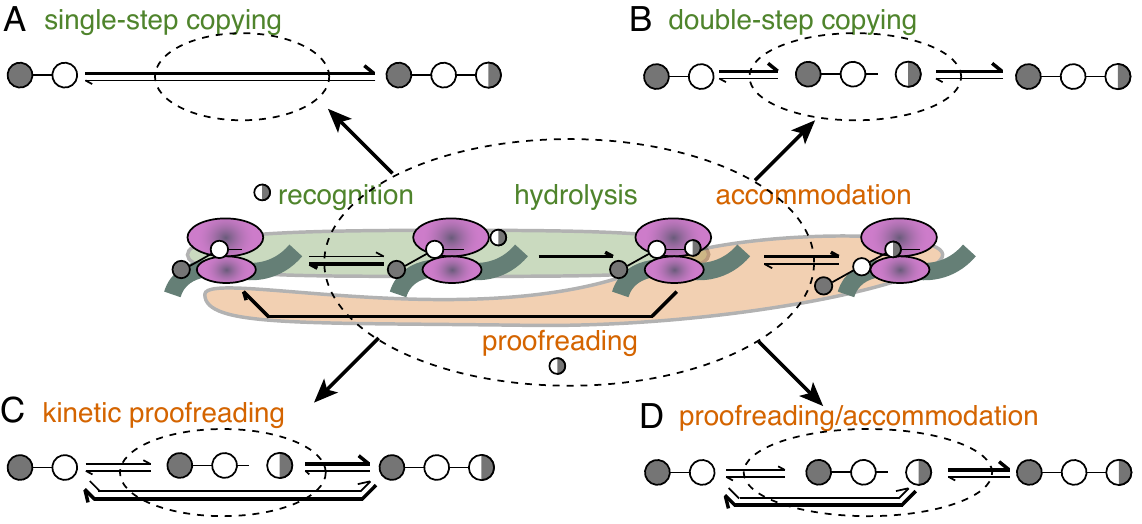}
\caption{{\bf Protocols of template-assisted polymerization}.  Panels
  A - D show the different copy protocols that we consider in this
  paper. A: single-step copolymerization without intermediate state
  (equivalent to Bennett's model \cite{bennett1979dissipation}. {\bf B.} double-step
  copying protocol. This can be seen as a variant of Bennett's model
  with the addition of an intermediate state. {\bf C.} kinetic proofreading
  and {\bf D.} proofreading/accommodation. 
  The central panel is a schematic of mRNA translation by the ribosome,  
  and illustrates how models A-D can be seen as building blocks of this 
  complex biological copying scheme, see e.g
  \cite{johansson2008kinetics,johansson2008rate,zaher2009fidelity}. In
  all panels, full gray and full white circles represent two different
  kinds of monomers. Half gray/half white circles represent monomers being either
  gray or white. 
    \label{fig:blocks}}
\end{center}
\end{figure}

   In this paper, we discuss how kinetic and energetic
  discrimination in different steps determine the minimum error in
  multi-step copy protocols. To this aim, we both present a systematic
  review of existing results and generalize some of them, for example
  by systematically studying all possible discriminating regimes of
  proofreading. We begin by reviewing the main features of Bennett's
  copolymerization model \cite{bennett1979dissipation} in the general
  case in which both energetic and kinetic discrimination are present
  \cite{sartori2013kinetic}, see Fig.~\ref{fig:blocks}A. We then
  generalize Bennett's model, which can be seen as the simplest
  possible copy protocol, \cite{bennett1979dissipation} to the
  biologically relevant cases of multi-step protocols (see also
  \cite{sartori2015}). In particular, our second copy protocol is made
  up of a linear chain of two states, Fig.~\ref{fig:blocks}B, while
  the third copy protocol, Fig.~\ref{fig:blocks}C, combines a linear
  chain of two states with a proofreading reaction acting from the
  last state. The last model, proofreading/accomodation, is a
  variation of model C in which proofreading acts from the
  intermediate state, see Fig.~\ref{fig:blocks}D. We shall see how
  this small variants changes quite dramatically the dependence of the
  error on the kinetic barriers of the three reactions. All these
protocols can be seen as ``building blocks'' of the mRNA translation
pathway, as represented in Fig.~\ref{fig:blocks}. By analytically
solving these models we find that, at given energy landscape, accurate
copying can be achieved only in a limited number of regimes, that we
fully characterize.

The paper is organized as follows. Section \ref{sec:model} introduces
our modeling framework, which has already been used in
\cite{sartori2015} to study the non-equilibrium thermodynamics of
copying. In section \ref{sec:results}, we present the
analytical solution for protocol A in Fig.~\ref{fig:blocks}. We study
in particular the different regimes in which error correction can be
achieved. In section \ref{sec:gensol}, we show how to solve more 
complex template-assisted polymerization models with intermediate
states. Sections \ref{sec:B}, \ref{sec:C}, and \ref{sec:D} present the
solution and analysis of the double-step copying, proofreading, and
proofreading/accommodation protocols respectively.  Section
\ref{sec:conclusions} is devoted to conclusions and future
perspectives.

\section{General model of template assisted polymerization}
\label{sec:model}

\begin{figure}
\begin{center}
\includegraphics[width=11.6cm]{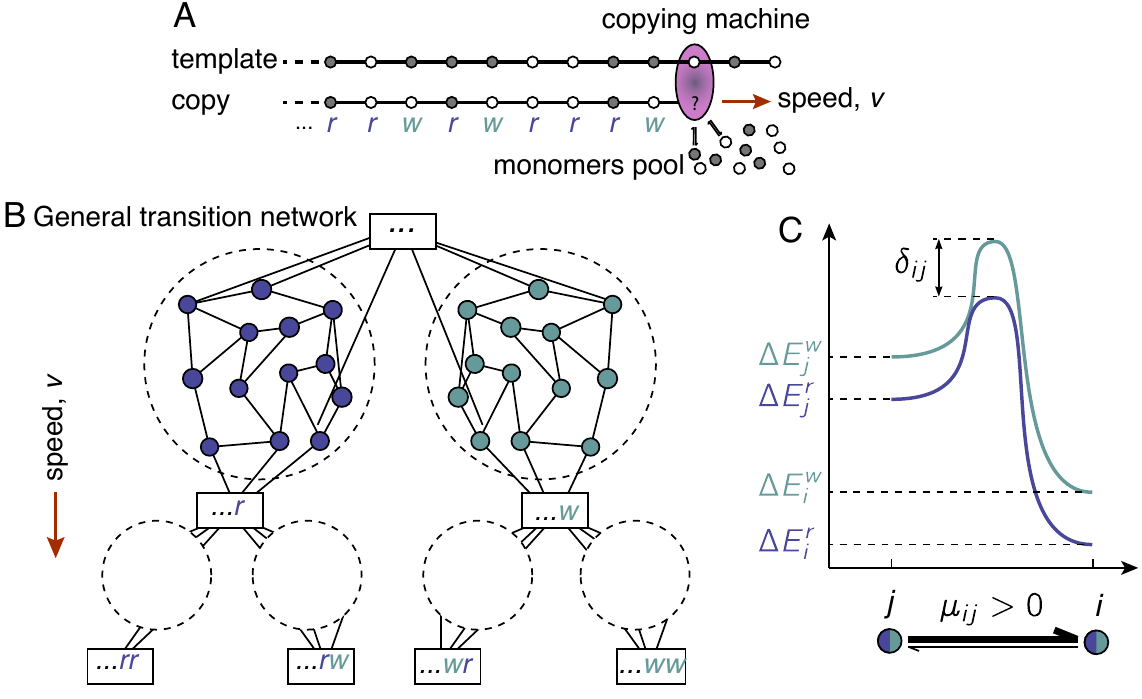}
\caption{{\bf Schematic summary of template-assisted polymerization.} {\bf A.} Sketch of the
  model in physical space. A molecular copying machine aids elongation 
  of a
  copy polymer, by trying to match candidate monomers with those on a
  pre-existing template polymer. {\bf B.} Transition network: rectangles
  represent the {\it main states}, and circles  the {\it intermediate states}
  leading to monomer incorporation. {\bf
    C.} Energy landscape corresponding to each transition, see
  Eq.~(\ref{eq_rates}). State $i$ can be either a main state
  (if $i=0$ or $i=n+1$) or an intermediate state (if $0<i\le n$) and the
  same for $j$.
\label{fig:scheme}}
\end{center} 
\end{figure}

Our modeling framework describes cyclic stochastic copying of a long
polymer. The system is made up of three parts: a pre-existing long
template polymer, which is not modified during the dynamics; a copy
polymer, which is assembled during the dynamics; and a copy machine,
which assists growth of the copy polymer and, in particular, tries to
match monomers incorporated in the copy with those in the template
(see Fig.~\ref{fig:scheme}A). For simplicity, we consider the monomer 
pool to be an infinite reservoir.

We focus on the case in which the physics does not
depend on the type of monomer, but only on whether it is a right match
$r$ (black-black or white-white in Fig.~\ref{fig:scheme}A) or a wrong
match $w$ (black-white or white-black).  The states of the system and the network of
possible stochastic transitions among them is then that in Fig.~\ref{fig:scheme}B.  After finalizing incorporation of a monomer, the
system is in a {\em main states}. Main states are represented in the
figure as rectangles with a string of $r/w$ inside. To process
a candidate monomer, the machine performs transitions through an
arbitrarily wired network of {\em intermediate states}. These
represent different conformational states of the copying machine, and allow for complex
error-correcting protocols. Intermediate states are represented as green and blue circles in
Fig.~\ref{fig:scheme}B, depending on whether the candidate monomer corresponds to a wrong or right match. Large dashed circles enclose each
sub-network leading to incorporation/rejection of the
candidate monomer.

Let us focus on a specific copy protocol, specified by the topology
of the network of its $n$ intermediate states. We denote with $\dots
rwr_i$ the $i$-th intermediate state leading to incorporation of a
right monomer and analogously for a wrong one, where the preceding
string of $r/w$ represents the already incorporated monomers. It is
convenient to identify, with a slight abuse of notation, $i,j=0$ with
the preceding main state and $i,j=n+1$ with the main state in which
the monomer is incorporated, i.e. $\dots rw r_0\equiv \dots rw $ and
$\dots rw r_{n+1}\equiv \dots rwr $. Irrespective of the existing $r/w$ string, transitions from state $j$ to
state $i$ occur at rate $k_{ij}^r$ and $k_{ij}^w$ for right and wrong
candidates respectively. To better understand the physics of the copying
process, it is useful to think that the stochastic transition rates
are determined by a) the energy of the states, b) chemical driving potentially fueling the different reactions, and c) activation
barriers separating the different states, as shown in
Fig.~\ref{fig:scheme}C. In terms of these quantities, the rates can be
expressed as

\begin{eqnarray}\label{eq_rates}
k_{ij}^r&=&w_{ij}\exp[(\Delta E^r_j+\mu_{ij}+\delta_{ij})/T]   \nonumber\\ 
k_{ij}^w&=&w_{ij}\exp[(\Delta E^w_j+\mu_{ij})/T]   \nonumber\\
k_{ji}^r&=&w_{ij}\exp[(\Delta E^r_i+\delta_{ij})/T]   \nonumber\\
k_{ji}^w&=&w_{ij}\exp[(\Delta E^w_i)/T]  .
\end{eqnarray}

In Eqs.~(\ref{eq_rates}), $\Delta E^r_j$ is the energy of right
state $j$, and similarly for wrong matches.  We take as
reference the initial main state, so that $\Delta E^w_0=\Delta
E^r_0=0$. $T$ is the temperature, measured from now on in units of the
Boltzmann constant ($k_B=1$). The parameter $\delta_{ij}$ accounts for
a difference in the activation barrier between state $j$ and $i$ for
right and wrong monomers. In particular, a positive value of
$\delta_{ij}$ means that the corresponding transition is kinetically
facilitated for right monomers with respect to wrong ones. The
chemical driving $\mu_{ij}>0$ favors the reaction $j\rightarrow i$,
irrespective of whether the match is right or wrong. Finally,
$\omega_{ij}$ is the inverse characteristic time of the transition
between states $j$ and $i$. The convention of Eqs.~(\ref{eq_rates})
differs slightly from that used in \cite{sartori2013kinetic} and, in
particular, simplifies picturing the energy landscape in multi-step
copy protocols.  Note that, in this case, the energies do not affect
only the backward rates but potentially also the forward rates and
thus kinetic discrimination.

The main focus of this paper is to study error-correcting regimes by
computing the error rate $\eta$ and speed $v$ for different specific copy
protocols. In all cases, we shall assume that the parameters
characterizing the difference between right and wrong monomers
(i.e. the $\Delta E^r_i$, $\Delta E^w_i$, and $\delta_{ij}$) are given, while the
chemical drivings $\mu_{ij}$ and the inverse timescales $\omega_{ij}$
can be tuned to optimize the error.

\section{Protocol A: single-step copolymerization}
\label{sec:results}
We begin by considering a protocol in which copying occurs without
intermediate states. Although this case has been studied in a number
of works (see
e.g. \cite{bennett1979dissipation,andrieux2008nonequilibrium,cady2009open,esposito2010extracting,sartori2013kinetic}),
it is useful to review the main results in order to introduce the
basic concepts that we will apply to more complex protocols.

The master equations governing this polymerization process can be written by
separating main states ending with a right incorporated monomer from
those ending with a wrong one:
\begin{align}
\label{eq:copol}
\dot{P}(\dots r) &= k^r_{10}P(\dots) +k^r_{01}P(\dots rr) + k^w_{01}P(\dots rw) \nonumber\\
  &-(k^r_{10}+k^w_{10}+k^r_{01}) P(\dots r)\quad, \nonumber\\
\dot{P}(\dots w) &= k^w_{10}P(\dots) +k^r_{01}P(\dots wr)+
k^w_{01} P(\dots ww )\nonumber\\
& -(k^r_{10}+k^w_{10}+k^w_{01}) P(\dots w) \quad .  
\end{align}

\begin{figure}
\begin{center}
\includegraphics{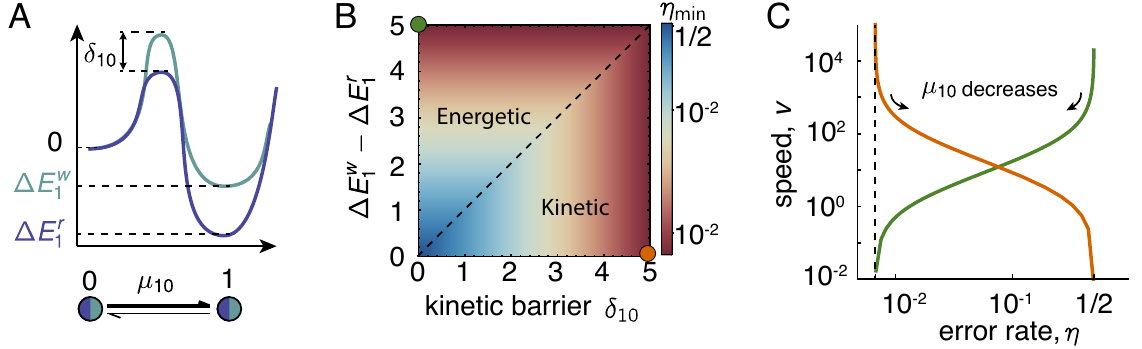}
\caption{{\bf Single-step protocol.} {\bf A.} Energy diagram . The dark
  blue curve corresponds to the landscape for a right match, and light
  blue for a wrong
  one. Below the states topology. {\bf B.} Heat map of the minimum
  error $\eta_{\rm min}$ as a function of the kinetic barrier
  $\delta_{10}$ and the energy difference $\Delta E^w_1-\Delta
  E^r_1$. A dashed line distinguishes the kinetic from the energetic
  regions. {\bf C.} Speed-error tradeoffs in the energetic and kinetic
  regions (see Eq.~\ref{speed_1} in the Appendix for the expression of
  the speed, and circles in panel B for values). The speed is measured
  in units of $\omega_{10}$.\label{fig_model1}}
\end{center}
\end{figure}

The system of equations (\ref{eq:copol}) can be solved at steady state
by using an {\em ansatz} of uncorrelated errors. This amounts to
assume that the steady-state probability of an arbitrary template-copy
sequence containing $N^r$ right matches and $N^w$ wrong matches is
given by
\begin{equation}\label{ansatz}
P(\dots)\propto(1-\eta)^{N^r} \eta^{N^w}\quad,
\end{equation}
where $\eta$ is the error rate to be determined {\em a
  posteriori}. Substituting (\ref{ansatz}) in the master equations
(\ref{eq:copol}) and imposing steady-state leads to a condition of the
form $\eta/(1-\eta)=v^w(\eta)/v^r(\eta)$, where
$v^w(\eta)=k_{10}^w-k_{01}^w\eta$ and
$v^r(\eta)=k_{10}^r-k_{01}^r(1-\eta)$ are the average incorporation
speeds of wrong and right monomers:  the stationary solution
  represents a non-equilibrium steady state characterized by a net
  growth speed (see e.g \cite{betterton2003motor}). Substituting the
expressions of the rates given by Eq.~(\ref{eq_rates}) and
Fig.~\ref{fig_model1}A finally yields
\begin{equation}\label{sol:model1}
\frac{\eta}{1-\eta}=
e^{-\delta_{10}/T}\frac{e^{\mu_{10}/T}-\eta e^{\Delta
    E_1^w/T}}{e^{\mu_{10}/T}-(1-\eta) e^{\Delta E_1^r/T}}\quad.
\end{equation}
Once the error is known, this speed is simply given by
$v=\eta v^w+(1-\eta)v^r$. By rearranging terms, Eq.~(\ref{sol:model1})
becomes a second-order equation for the error $\eta$ that can be
easily solved.  It is instructive to discuss the properties of this
solution in some depth and, in particular, how varying the chemical
driving $\mu_{10}$ affects the error rate.

A copy is performed quasi-statically when the growth speed vanishes,
$v=0$. This requires that $v^w=$ and $v^r=0$, i.e. both numerator and
denominator of Eq.~(\ref{sol:model1}) vanish.  Imposing this
condition, one finds that the error reaches its equilibrium value
$\eta_{\rm eq}=1/(1+e^{(\Delta E_1^w-\Delta E_1^r/T})$. To reach this
limit, the chemical driving $\mu_{10}$ must be tuned to its stall
value $\mu_{\rm st}\equiv -T\log(e^{-\Delta E_1^w/T}+e^{-\Delta
  E_1^r/T})$. As originally observed by Bennett, the stall chemical
driving is {\em negative}, since to arrest polymerization it needs to
oppose the entropic force driving growth of the chain
\cite{bennett1979dissipation,andrieux2008nonequilibrium,esposito2010extracting,sartori2013kinetic}. 
This entropic force can be physically understood by thinking that the
number of possible states grows
exponentially as the copy polymer elongates.
In the following, we shall focus only on the case in which the copying
speed is non-negative, $\mu_{10}\ge \mu_{\rm st}$ (see
\cite{andrieux2013information} for considerations on the case of
negative speed).

Upon increasing the chemical driving from its stall value, the speed
$v$ increases monotonically. The error approaches, also monotonically,
its large-driving irreversible limit $\eta_{\rm
  irr}=e^{-\delta_{10}/T}/(1+e^{-\delta_{10}/T})$. This implies that
the error is an increasing or decreasing function of the chemical
driving depending on whether $\eta_{\rm irr}$ is larger or smaller
than $\eta_{\rm eq}$, respectively.

This finding is summarized in Fig.~\ref{fig_model1}B, where we
represent the minimum possible error $\eta_{\rm min}$ as a function of the energy
difference $\Delta E^w_1-\Delta E^r_1$ and the kinetic barrier $\delta_{10}$. Depending
on which is the larger between these two quantities, we can distinguish
between two different regions in parameter space:

\begin{itemize}
\item {\bf Kinetic region.} This region is defined by the condition
  $\delta_{10}>\Delta E^w_1 -\Delta E^r_1$ in
  Fig.~\ref{fig_model1}B, which implies $\eta_{\rm irr}<\eta_{\rm eq}$.  In this region, the minimum error is equal to
  $\eta_{\rm irr}$, and is achieved in the limit of a very large chemical driving
$\mu_{10}\gg\mu_{\rm st}$.
\item {\bf Energetic region.} This is the region
  $\delta_{10} < \Delta E^w_1 - \Delta E^r_1$. The minimum error is the equilibrium
  error $\eta_{\rm eq}$, which depends only on the energy difference
  $\Delta E^w_1-\Delta E^r_1$. The minimal error is 
  achieved when the chemical driving tends to its stall value
  $\mu_{\rm st}$.
\end{itemize}

Symmetric points on different sides of the diagonal in
Fig.~\ref{fig_model1}B  are characterized
by the same minimum error. However, being in one or the other region
affects the way the system responds to a change in the chemical
driving. On the other hand, increasing the chemical driving always
increases the copying speed.
This means that the error is a decreasing function of the chemical
driving and the copying speed in
the kinetic region, and an increasing function in the energetic
region. These two opposite tradeoffs are shown in Fig.~~\ref{fig_model1}C.

\section{Solution to a general protocol}\label{sec:gensol}

The approach used in the previous section can be generalized
for arbitrarily complex models \cite{sartori2015}.
We now discuss how to find the steady-state solution for a general
protocol including intermediate states. 

 We start by writing the
general master equations. It is convenient to define the probability
fluxes $\mathcal{J}^r_{ij}(\dots)=k^r_{ij}P(\dots r_j)-k^r_{ji}P(\dots
r_i)$ between pairs of states.  The master equations for the
intermediate states then read
\begin{align}
\label{eq:intermediate}
\dot{P}(\dots r_i)=\sum\limits_{j=0}^{n+1} \mathcal{J}^r_{ij}(\dots)\quad {\rm a
nd}\quad
\dot{P}(\dots w_i)=\sum\limits_{j=0}^{n+1}\mathcal{J}^w_{ij}(\dots)\quad ,
\end{align}
while the master equations for the main states are
\begin{align}\label{eq:main}
  \dot{P}(\dots w)&=\sum\limits_{j=0}^{n+1}\left[ \mathcal{J}^w_{n+1j}(\dots)
    -\mathcal{J}^r_{j0}(\dots w)-\mathcal{J}^w_{j0}(\dots w)\right]\quad,\nonumber\\
  \dot{P}(\dots
  r)&=\sum\limits_{j=0}^{n+1}\left[\mathcal{J}^r_{n+1j}(\dots)
    -\mathcal{J}^r_{j0}(\dots r)-\mathcal{J}^w_{j0}(\dots r)\right] \quad.
\end{align}
Note that Eqs.~(\ref{eq:main}) correctly reduce to
Eqs.~(\ref{eq:copol}) in the case of
$n=0$.

The general system of equations above can be solved at steady state
for any choice of the copy protocol. A solution can be obtained also
in this case by assuming uncorrelated errors, as in
Eq.~(\ref{ansatz}).  However, in this case, we have to further assume 
that steady-state probabilities of intermediate states are
proportional to the probability of the main state before attempted
monomer incorporation:
\begin{equation}\label{ansatz2}
P(\dots r_i)=P(\dots) p_i^r\quad,
\end{equation}
where we introduced the {\it occupancies} $p_i^r$, and similarly for
wrong monomers. Note that Eq.~(\ref{ansatz}) implies $p_0^r=p_0^w=1$,
$p_{n+1}^r=(1-\eta)$ and $p_{n+1}^w=\eta$.  

For any given network of intermediate states, one can impose the
steady-state condition and the solution {\it ansatz} in
Eqs.~(\ref{ansatz}) and (\ref{ansatz2}).  Doing so,
Eq.~(\ref{eq:intermediate}) becomes a linear system of equations for
the occupancies of the intermediate states. Solving this linear
  systems yields explicit expressions for the occupancies in terms of
  the transition rates and the yet unknown error rate $\eta$.
  Substituting the occupancies into Eqs.~(\ref{eq:main}) results into
  equations formally identical to Eqs.~(\ref{eq:copol}), with
  ``renormalized'' rates of incorporation/removal of right/wrong
  monomers that take into account the effect of the intermediate
  states.  In short, after having performed steady-state elimination
  of the intermediate states, any arbitrarily complex
  template-assisted polymerization protocol can be formally reduced to
  the simple model without intermediate states described in section
  \ref{sec:results}. As a consequence, the equation for the error
rate $\eta$ can be always put in the form
\begin{equation}
\label{eq:error}
\frac{\eta}{1-\eta}=\frac{v^w(\eta)}{v^r(\eta)} \quad ,
\end{equation}
where,  thanks to the steady-state elimination of the
  occupancies, $v^w(\eta)$ and $v^r(\eta)$ can be explicitly
expressed in terms of the rates as in the case of
  Eq. (\ref{sol:model1}). Also in this case, Eq. \ref{eq:error} is an
  algebraic equation for the only unknown $\eta$ and its solution
  completes the solution of the model. We stress once more that,
while in the following we will focus on simple protocols with only one
intermediate state, the solution strategy described in this section is
general: the difficulties of solving larger networks of intermediate
states are only of algebraic nature.

\section{Protocol B: double-step copying}
\label{sec:B}

The second-simplest possible copy protocol is a linear chain of
states where final monomer accommodation is preceded by an intermediate
state, see Fig.~\ref{fig:blocks}. In the most general case both steps can discriminate kinetically and energetically, and the energy landscape is that shown
in Fig.~\ref{fig_doublecopying}A. Following the general solution
strategy described in previous subsection we can determine the error rate $\eta$ and speed $v$ of this complex copying pathway. Solving at steady-state
Eqs.~(\ref{eq:intermediate}) gives the occupancies of the
intermediate states,
$p_1^r=[k_{10}^r+(1-\eta)k_{12}^r]/(k_{01}^r+k_{21}^r)$ and
$p_1^w=(k_{10}^w+\eta k_{12}^w)/(k_{01}^w+k_{21}^w)$. Using further the expressions for $v^w$ and $v^r$ (see Appendix), the equation for
the error rate can be written as
\begin{equation}
\frac{\eta}{1-\eta}=\frac{p_1^w k_{21}^w-k_{12}^w\eta}{p_1^r
  k_{21}^r-k_{12}^r(1-\eta)} \quad.
\end{equation}
Substituting the expressions of the occupancies as well as the parametrization of the rates yields
\begin{equation}\label{sol:model2}
\frac{\eta}{1-\eta}=e^{-\frac{(\delta_{10}+\delta_{21})}{T}}\frac{\left[e^{\frac{\delta_{10}}{T}}+re^{\frac{\delta_{21}+\mu_{21}}{T}}\right]}{\left(1+r
  e^{\frac{\mu_{21}}{T}}\right)}\frac{\left(e^{\frac{\mu_{10}+\mu_{21}}{T}}
  - \eta e^{\frac{\Delta E_2^w}{T}}
\right)}{\left[e^{\frac{\mu_{10}+\mu_{21}}{T}} 
-(1-\eta) e^{\frac{\Delta E_2^r}{T}}\right]}
\end{equation}
where we introduced the dimensionless ratio of time scales of the two
reactions $r=\omega_{21}/\omega_{10}$. 

The above equation allows to calculate the error rate $\eta$ given the
discrimination barriers $\delta_{10}$ and $\delta_{21}$, the energy differences $\Delta E_2^w$ and $\Delta E_2^r$, the chemical drivings $\mu_{10}$ and $\mu_{21}$, and the ratio of rates $r$. In
this case, optimizing the error at fixed energy landscape and discrimination barriers is more
subtle, as there are three free parameters in the system: the
two chemical drivings and the time scale
ratio. By taking different limits, we can identify three regions
in parameter space:

\begin{figure}
\begin{center}
\includegraphics{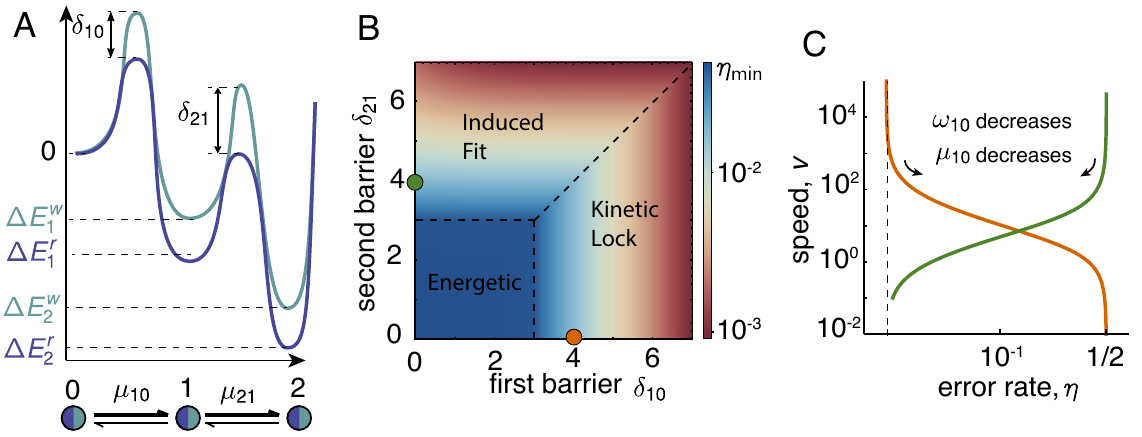}
\caption{{\bf Double-step copying}. {\bf A.} Energy landscape showing 
kinetic and energetic differences in both steps. {\bf B.}
  Minimum error and regions. Parameters are $\Delta E_2^r=\Delta
  E_1^r=0$, $\Delta E_2^w=\Delta E_1^w=3T$. The
  chemical drivings and the relative time scales of the different
  reactions have been determined by numerically minimizing the
  error. {\bf C.} Speed-error tradeoff for points in the Induced fit and kinetic lock regimes, see B. Speed is in units of $\omega_{10}$. \label{fig_doublecopying}}
\end{center}
\end{figure}

\begin{itemize}

\item {\bf Energetic region.} This region is defined by $\Delta
  E_2^w-\Delta E_2^r>\delta_{10}$ and $\Delta E_2^w-\Delta
  E_2^r>\delta_{21}$. In this region, the equilibrium error $\eta_{\rm
    eq}=e^{-\Delta E_2^w/T}/(e^{-\Delta E_2^w/T}+e^{\Delta E_2^r/T})$
  is the minimal error of the system. It is approached on the
  quasi-static limit, as can be derived from Eq.~(\ref{sol:model2}) by
  taking the limit of vanishing speed. Also in this case, to
  achieve equilibrium, the total chemical driving $\mu_{10}+\mu_{21}$
  must be negative and equal to the stall value $\mu_{\rm st}$.

\item {\bf Kinetic Lock.} This is the region $\delta_{10}>
  \Delta E_2^w-\Delta E_2^r$ and $\delta_{10}>\delta_{21}$.  The minimum
  error is equal to $\eta_{\rm irr}$ defined in the previous
  section. It is achieved in a strongly-driven regime in which
  $\mu_{10}, \mu_{21}\rightarrow \infty$.  Note that, since the
  dynamics is completely irreversible, only the first barrier
  influences the error. The system is thus kinetically locked after
  the first step. 

\item {\bf Induced fit.} In this region the parameters satisfy
  $\delta_{21}> \Delta E_2^w-\Delta E_2^r$ and $\delta_{21}>
  \delta_{10}$. The minimum error rate is $\eta\approx
  e^{-\delta_{21}/T}$. It can be achieved by taking first the limit
  $r\rightarrow 0$ and then $\mu_{10}, \mu_{21}\rightarrow\infty$. We
  term this regime ``induced fit'' as it is reminiscent of the
  mechanism in enzyme kinetics where discrimination occurs via a
  kinetic mechanism acting after initial binding, see e.g. \cite{pape1999induced}.
\end{itemize}

Summarizing, copies can be produced near equilibrium, by using an
energy difference; or out of equilibrium, by using either the first or
the second kinetic barrier. Thus, also in the double-step copying 
protocol energetic and kinetic discrimination strategies can not be combined. Moreover, it is impossible to exploit both kinetic barriers: in
particular, if the first reaction is strongly driven, no mechanism
acting afterwards can alter the overall accuracy.

These three regions are shown in Fig.~\ref{fig_doublecopying}B,
where we numerically minimized the error at fixed values of the
energies parametrically varying the two barrier differences $\delta_{10}$ and $\delta_{21}$. Finally, due to
the complex dependence of the error on parameters, also in this case
one can find negative and positive speed-error relations depending on
which parameter is altered to tune the error and in which region the
model is operating, see example in Fig.~\ref{fig_doublecopying}C.

\section{Protocol C: Kinetic Proofreading}
\label{sec:C}

\begin{figure}
\begin{center}
\includegraphics{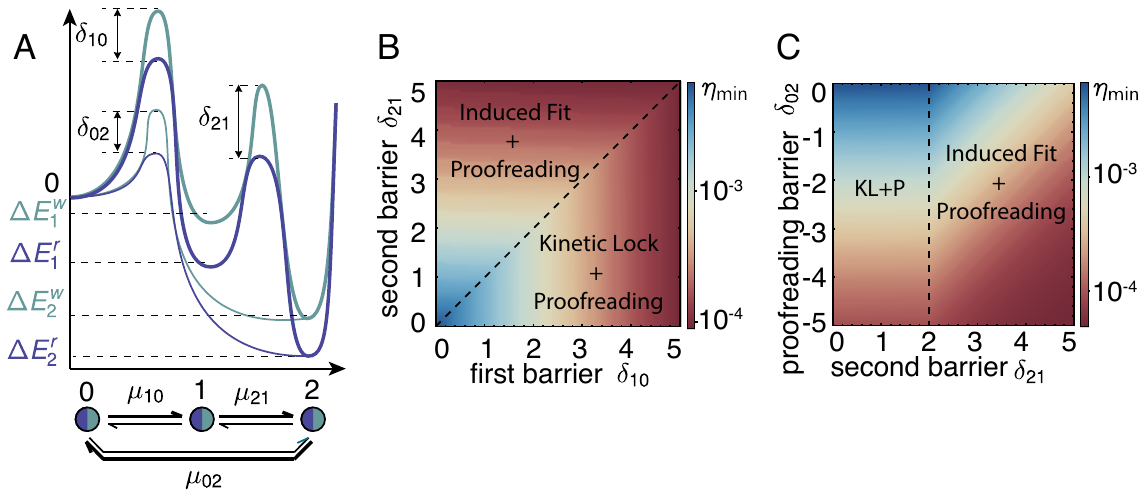}
\caption{{\bf Kinetic Proofreading.} {\bf A}: Energy landscape showing a generic double copy landscape together with the proofreading reaction which discriminates kinetically through the barrier $\delta_{02}<0$. {\bf B.} \& {\bf
    C.} Regions and minimum error. Parameters are $\Delta E_2^r=\Delta E_1^r=0$,
  $\Delta E_2^w=\Delta E_1^w=3T$.  In panel B, we fixed $\delta_{02}=-2T$ while in
  panel C we fixed $\delta_{10}=2T$. The chemical drivings and the
  relative time scales of the different reactions have been determined
  by numerically minimizing the error. \label{fig_proofreading}}
\end{center}
\end{figure}

We now discuss a proofreading protocol in the spirit of Hopfield's model,
where a double-step copying scheme is assisted by a proofreading
reaction $2\to0$ which is driven backward and tries to correct wrongly copied
monomers, see Fig.~\ref{fig:blocks} and the general energy landscape in
Fig.~\ref{fig_proofreading}A. Proceeding as before, we obtain the
equation for the error rate
\begin{equation}
\frac{\eta}{1-\eta}=\frac{p_1^w k_{21}^w+k_{20}^w-(k_{12}^w+k_{02}^w)\eta}{p_1^r
  k_{21}^r+k_{20}^r-(k_{12}^r+k_{02}^w)(1-\eta)} \quad,
\end{equation}
where $p_1^w$ and $p_1^r$ have the same expression as in the previous
section.  Substituting the formula of the rates yields a lengthy
expression (see Appendix).
To simplify it, we exploit the fact that, in an efficient
error-correction scheme, the second step and the proofreading
reaction are irreversible, so that $k_{12}^r$, $k_{12}^w$, $k_{20}^r$,
and $k_{20}^w$ can be neglected. These assumptions are the same used to simplify
the reaction scheme in Hopfield's original model
\cite{hopfield1974kinetic}. In this approximation, the error satisfies
\begin{equation}\label{proof2}
\frac{\eta}{1-\eta}=\frac{\left[e^{\frac{\delta_{10}}{T}}+re^{\frac{\delta_{21}+\mu_{21}}{T}}\right]}{\left(1+r
  e^{\frac{\mu_{21}}{T}}\right)} \frac{K-\eta e^{\frac{\Delta E_2^w}{T}}(1+r)}{K
  e^{\frac{\delta_{21}+\delta_{10}}{T}} -(1-\eta)
  e^{\frac{\Delta E_2^r+\delta_{02}}{T}}\left(  e^{\frac{\delta_{10}}{T}}+r e^{\frac{\delta_{21}}{T}}  \right) }\;,
\end{equation}
where we defined  $r=\omega_{21}/\omega_{10}$ as previously and $K=(\omega_{21}/\omega_{02})
e^{\frac{\mu_{10}+\mu_{21}-\mu_{02}}{T}}$. Increasing the proofreading
driving $\mu_{02}$ (i.e. decreasing $K$) improves the accuracy as more and more
copies are proofread. However, this also reduces the copying speed;
the limiting error is then obtained in the limit of vanishing
speed. Determining $K$ in this way and substituting back in the
numerator of Eq.~(\ref{proof2}) results in a general expression for the minimal error
\begin{equation}\label{proof_err}
\eta_{\rm min}(v\to0)\approx
e^{\frac{\Delta E^r_2-\Delta E^w_2+\delta_{02}}{T}}\frac{\left(e^{\frac{\delta_{10}}{T}}+r
    e^{\frac{\delta_{21}}{T}}\right)}{(1+r)
  e^{\frac{\delta_{21}+\delta_{10}}{T}}} \quad,
\end{equation}
where the irreversible assumption explicitly excludes the trivial energetic region. But even with this assumption, the expression above is richer than that of the irreversible error $\eta_{\rm irr}$  in previous sections.

The first factor in Eq.~(\ref{proof_err}) represents the 
discrimination due to proofreading, and includes the energy difference $\Delta E^r_2-\Delta E^w_2<0$ and the kinetic barrier $\delta_{02}<0$. The following fraction represents the 
discrimination of the copying
reaction, and through its dependence on the ratio $r$, different operating regions can be distinguished. As
Eq.~(\ref{proof_err}) is always monotonic in $r$, we can identify two
different regions depending on whether the error decreases or
increases with $r$:

\begin{itemize}
\item {\bf Kinetic Lock + Proofreading.} This is the region in which
  $\delta_{10}>\delta_{21}$ The minimum error is obtained in the limit
  $r\gg1$ of Eq.~(\ref{proof_err}), so that the discrimination factor
  of the copying reaction, i.e. the fraction in Eq.~(\ref{proof_err}),
  tends to $\exp(-\delta_{10}/T)$.
\item {\bf Induced Fit + Proofreading.} In this region the parameters satisfy
  $\delta_{21}>\delta_{10}$. To obtain
  the minimum error, one should take the limit $r\ll1$ of
  Eq.~(\ref{proof_err}), so that the  fraction in Eq.~(\ref{proof_err})
   tends to $\exp(-\delta_{21}/T)$. 
\end{itemize}

The existence of these two regions is shown in
Fig.~\ref{fig_proofreading}B and \ref{fig_proofreading}C, where we
plot the minimum error as a function of the kinetic barrier of the second copy
reaction $\delta_{21}$ and the proofreading barrier
$\delta_{02}$. According to Eq.~{\ref{proof_err}}, the minimum error
always decreases exponentially with $\delta_{02}$, consistently with
the figure. On the other hand, the minimum error depends on
the maximum between $\delta_{10}$ and $\delta_{21}$, so that the error
in the figure depends on $\delta_{21}$ only when $\delta_{21}>\delta_{10}$.

This protocol is a generalization of previous proofreading models as those proposed by Hopfield
 \cite{hopfield1974kinetic} and Bennett \cite{bennett1979dissipation} .
In particular, Hopfield's model corresponds to the choice
of $\delta_{10}=\delta_{02}=0$, $\Delta E_2^w=\Delta E_1^w$ and
$\Delta E_2^w=\Delta E_1^r$. Moreover, as the forward rates of the second reaction
are equal for right and wrong incorporations, one should fix
$\delta_{21}=\Delta E_2^w-\Delta E_2^r$. With this choice, the minimum error in
Eq.~(\ref{proof_err}) correctly becomes $\eta_{\rm min}\approx
\exp[-2(\Delta E_2^w-\Delta E_2^r)/T]$, i.e. the square of the equilibrium error. On
the other hand, Bennett's model of proofreading
can be seen as the case in which
accuracy in copying is achieved via kinetic lock with no energetic differences, $E_2^w=E_2^r$. In this case we have $\eta_{\rm min}\approx\exp[-(\delta_{21}-\delta_{02})/T]$.

\section{Protocol D: Proofreading/accommodation}
\label{sec:D}

\begin{figure}
\begin{center}
\includegraphics{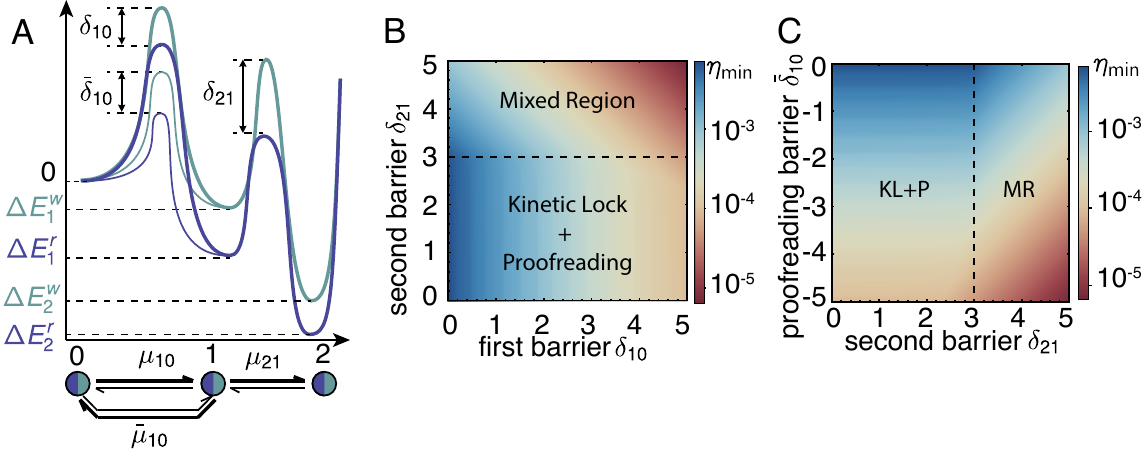}
\caption{{\bf Proofreading/accommodation}. {\bf A} Energy
  landscape  {\bf B.} \&
  {\bf C.} Regions and minimum error.  Parameters are $\Delta E_2^r=\Delta E_1^r=0$,
  $\Delta E_2^w=\Delta E_1^w=3T$,
  $\delta_{01}=2T$. The chemical drivings and the relative time scales
  of the different reactions have been determined by numerically
  minimizing the error. \label{fig_proofaccom}}
\end{center}
\end{figure}

In this protocol, inspired by the later stage of the mRNA translation
pathway (see Fig.~\ref{fig:blocks}), proofreading acts from the
intermediate step of the reaction. The energy landscape is shown
in Fig.~\ref{fig_proofaccom}A.  Notice that in this case there are
two sets of rates connecting state $0$ with state $1$, corresponding
to either the first copying step or the proofreading reaction. To
resolve this ambiguity, we call $k^r_{10}$ the binding rate of right
monomer for the copying reaction and $\bar{k}^r_{10}$ the
corresponding rate of the proofreading reaction, and similarly for the
other rates. We adopt the same notation for the energies. The
occupancies of the intermediate states then are
$p_1^r=[k_{10}^r+\bar{k}_{10}^r+(1-\eta)k_{12}^r]/(k_{01}^r+\bar{k}_{01}^r+k_{21}^r)$
and $p_1^w=(k_{10}^w+\bar{k}_{10}^w+\eta
k_{12}^w)/(k_{01}^w+\bar{k}_{01}^w +k_{21}^w)$.  The solution in terms
of the occupancies is the same as in the double-copying model
\begin{equation}
\frac{\eta}{1-\eta}=\frac{p_1^w k_{21}^w-k_{12}^w\eta}{p_1^r
  k_{21}^r-k_{12}^r(1-\eta)}\quad.
\end{equation}

Substituting the expressions of the rates results in
Eq.~(\ref{sol:model4}) in the Appendix. If we assume that all drivings
are large, and that proofreading compensates copying so that the net
speed vanishes as before, we obtain from the previous expression
that the minimum error is
\begin{equation}\label{minKL4}
\eta_{\rm min}(v\rightarrow 0)\approx e^{\frac{-\delta_{10}+E^r_2-E^w_2+\bar{\delta}_{10}}{T}}\quad,
\end{equation}
i.e. the minimum error associated to the first barrier, times the
proofreading error reduction factor.  To exploit the second barrier, we
should take the opposite limit of large speed. In this case, the
error tends to
\begin{equation}\label{min_fast4}
\eta_{\rm min}(v\rightarrow \infty)\approx
e^{\frac{-\delta_{10}-\delta_{21}}{T}}\frac{\left[e^{\frac{\delta_{10}}{T}}
+\bar{r}e^{\frac{\bar{\delta}_{10}}{T}}+re^{\frac{\delta_{21}+\mu_{21}}{T}}\right]}
{\left(1+\bar{r}+r e^{\frac{\mu_{21}}{T}}\right)}
\end{equation}
with $\bar{r}=\bar{\omega}_{01}/\omega_{10}$. Assuming, as before, that
the proofreading barrier is non-positive $\bar{\delta}_{10}\le0$, that
the other two barriers are non-negative and that $\Delta E^w-\Delta
E^r\ge 0$, the minimum error is found in one of the following two
regions:
\begin{itemize}
\item {\bf Kinetic Lock + Proofreading.} This region is defined by the
  condition $\delta_{21}< \Delta E_2^w-\Delta E_2^r$ and is analogous to the one in
  the proofreading model of the previous section. The expression for the minimum error is thus
  given by Eq.~(\ref{minKL4}).
\item {\bf Mixed Region.} In this region, the parameters satisfy
  $\delta_{21}> \Delta E_2^w-\Delta E_2^r$. The minimum error is
  obtained by taking the limit $\bar{r}\gg r$ and $\bar{r}\gg 1$ in
  Eq.  (\ref{min_fast4}) and is equal to $\eta_{\rm
    min}=\exp[(\bar{\delta}_{10}-\delta_{10}-\delta_{21})/T]$. As
  proofreading is performed from the intermediate state, in this model
  it is possible to exploit all three kinetic barriers at the same
  time. This is at variance with models B and C where only one
  barrier, either $\delta_{10}$ or $\delta_{21}$, affects the minimum
  error in each region.
\end{itemize}

%It can be easily shown that other limits of
%Eq.~(\ref{min_fast4}) yield minimum errors which are alway worse than these
%two values. 
%

Also in this case we confirmed the existence of these two regions by
numerically minimizing the error for different values of the
discrimination parameters. The results are summarized in
Fig.~\ref{fig_proofaccom}B and \ref{fig_proofaccom}C.

\section{Conclusions}\label{sec:conclusions}

In this paper, we considered four copy protocols inspired by
the building blocks of the mRNA translation pathway. We have shown
that, despite the relatively large number of parameters, their copying
accuracy can be analyzed in full generality.  Our analysis
  reveals that each of these copy protocols can function in several
  operating regimes, which generalize of the kinetic and energetic
  regimes of the single-step copy protocol (see
  Fig.~\ref{fig:potential}). It remains an open problem to find
  generic rules by which kinetic and energetic discriminating steps
  can be combined in an arbitrarily complex copying protocol.

As for 
the single-step copy, the regimes of a complex copy protocol are also 
characterized by a distinctive minimal error
and a strategy to achieve it. Each regime can be associated to
particular tradeoffs between chemical driving, speed, and error. 
It interesting to explore how these trade-offs are resolved in actual
biological systems. For example, it has been argued that
initial energetic differences that could allow for energetic
discrimination if equilibrated are not fully exploited in translation
\cite{thompson1982accuracy,gromadski2004kinetic,johansson2008rate,johansson2008kinetics}. This
constitutes an example where, in the presence of a positive
error-speed tradeoff, optimizing the speed seems to be more
relevant.

In this paper, we focused on optimizing the error rate alone, and found
that it can be minimized in different regimes corresponding to very
different values of the speed.  One might also
wonder whether the error rate is determined
by some general optimization principle. Some ideas in this
direction have been recently explored.  By looking at the first two
steps of translation, Savir and Tlusty \cite{savir2013ribosome}
concluded that the energy landscape is sculpted to optimize decoding
of information.  More recently, Rao and Peliti proposed a measure of
optimality of copying reaction which includes a cost for dissipated work 
\cite{rao2015thermodynamics}.

The modeling framework discussed in this paper is very general.  Given
the possibility of analytically solving any copying protocol, as
discussed in section \ref{sec:gensol}, it constitutes a natural
starting point to study the dynamics of many complex copying
machines. A possible further generalization would be to include the
possibility of performing stochastic transitions {\em among} different
sub-networks.  This possibility is important to study 
  backtracking, i.e. the possibility of copying enzymes to
error-correct entire strings of the copy polymer containing a high error
fraction, see
e.g. \cite{galburt2007backtracking,voliotis2008fluctuations,mellenius2015dna}.

From the point of view of building up a general theory
\cite{parrondo2015thermodynamics}, it would also be interesting to
understand the relationship between copying and other
information-processing tasks performed by biological systems. For
example, an interesting analogy exists with sensing, where cells have
to detect a (noisy) external signal and internally store its value
\cite{lan2012energy,bo2015thermodynamic}. Some aspects of this analogy have already been
explored, for example with the concept of kinetic proofreading
\cite{hartich2015nonequilibrium}. It is however possible that this
analogy can be pushed further. For example, thinking in terms of
kinetic and energetic discrimination, as discussed in this paper,
could help understanding regimes of sensing and other
information-processing tasks.

\begin{acknowledgements}
  This work was supported by the Ministerio de Economia y
  Competividad (Spain) and FEDER (European Union), under project
  FIS2012-37655-C02-01.
\end{acknowledgements}

\section*{Appendix: exact expressions}

In this Appendix, we report the exact expression for the solutions of
the proofreading models, that have been omitted or simplified in the Results
section for ease of reading. 

\subsection*{Template-assisted polymerization without intermediate states}
The exact equation for the error is in the Results section.  The
copying speed is equal to
\begin{equation}\label{speed_1}
v=k_{10}^w+k_{10}^r-k_{01}^w\eta-k_{01}^r(1-\eta).
\end{equation}
\subsection*{Double-step copying}

Also in this case, the exact equation for the error is in the Results
section.  The expression for the speed is
\begin{equation}\label{speed_2}
v=\mathcal{N}[p_1^w k_{21}^w+p_1^r
  k_{21}^r-k_{12}^w\eta-k_{12}^r(1-\eta) ]
\end{equation}
where we introduced the normalization factor for the occupancies $\mathcal{N}$,
which in this case is equal to $\mathcal{N}=(1+p_1^w+p_1^r)^{-1}$.
\subsection*{Kinetic proofreading}

The error rate of the proofreading model satisfies the equation
\begin{eqnarray}
\frac{\eta}{1-\eta}&=&\frac{\left[e^{\frac{\delta_{10}}{T}}+re^{\frac{\delta_{21}+\mu_{21}}{T}}\right]}{\left(1+r
  e^{\frac{\mu_{21}}{T}}\right)}\times \\
&\times&\frac{r e^{\frac{\mu_{10}+\mu_{21}}{T}}+r_p\left(1+r
    e^{\frac{\mu_{21}}{T}}\right)-\eta  e^{\frac{\Delta E_2^w}{T}} \left[r
   +(1+r) r_p
  e^{\frac{\mu_{02}}{T}}\right]}{r
e^{\frac{\mu_{10}+\mu_{21}+\delta_{21}+\delta_{10}}{T}}+r_p e^{\frac{\delta_{20}}{T}}\left(e^{\frac{\delta_{10}}{T}}+r
    e^{\frac{\mu_{21}+\delta_{21}}{T}}\right)-(1-\eta)e^{\frac{\Delta E_2^r}{T}}A} \nonumber
\end{eqnarray}
where $r_p=\omega_{02}/\omega_{10}$ and $A=\left(r e^{\frac{\delta_{10}+\delta_{21}}{T}}
   +r_p e^{\frac{\delta_{10}+\delta_{02}+\mu_{02}}{T}} +r r_p e^{\frac{\delta_{21}+\delta_{02}+\mu_{02}}{T}} \right) $.

The copying speed is equal to
\begin{equation}\label{speed_3}
v=\mathcal{N}[p_1^w k_{21}^w+k_{20}^w + p_1^r
  k_{21}^r+k_{20}^r-(k_{12}^w+k_{02}^w)\eta -(k_{12}^r+k_{02}^w)(1-\eta)]
\end{equation}
where $\mathcal{N}$ is defined as previously.
\subsection*{Proofreading/accommodation}

The error rate satisfies

\begin{eqnarray}\label{sol:model4}
\frac{\eta}{1-\eta}=e^{-\frac{\delta_{21}}{T}}\frac{\left[e^{\frac{\delta_{10}}{T}}+\bar{r}e^{\frac{\bar{\delta}_{10}}{T}}+re^{\frac{\delta_{21}+\mu_{21}}{T}}\right]}
{\left(1+\bar{r}+r e^{\frac{\mu_{21}}{T}}\right)}\times \nonumber\\
\times
\frac{e^{\frac{\mu_{21}}{T}}\left(e^{\frac{\mu_{10}}{T}}+\bar{r}\right)-\eta
  e^{\frac{\Delta E_2^w}{T}}(1+\bar{r}e^{\frac{\bar{\mu}_{10}}{T}})   }
{e^{\frac{\mu_{21}}{T}}\left(e^{\frac{\mu_{10}+\delta_{10}}{T}}+\bar{r}e^{\frac{\bar{\delta}_{10}}{T}}\right)-(1-\eta)
  e^{\frac{\Delta E_2^r}{T}}
\left(e^{\frac{\delta_{10}}{T}}+\bar{r}e^{\frac{\bar{\mu}_{10}+\bar{\delta}_{10}}{T}}\right)}
%\frac{\left(e^{\frac{\mu_{10}+\mu_{21}}{T}}
 % - \eta e^{\frac{\Delta E_2^w}{T}}
%\right)}{\left[e^{\frac{\mu_{10}+\mu_{21}}{T}} 
%-(1-\eta) e^{\frac{\Delta E_2^r}{T}}\right]}
\end{eqnarray}

and the copying speed is 
\begin{equation}\label{speed_4}
v=\mathcal{N}[p_1^w k_{21}^w+p_1^r
  k_{21}^r-k_{12}^w\eta-k_{12}^r(1-\eta)].
\end{equation}

\bibliographystyle{spmpsci}      % mathematics and physical sciences

\bibliography{regimes_jsp}   % name your BibTeX data base

\end{document}